\documentclass[preprint,pre,aps,showpacs]{revtex4}
\usepackage{graphicx,color}
\begin{document}

\title{Symmetry energy from fragment observables in the canonical
thermodynamic model} 

\author{G. Chaudhuri$^1$, F. Gulminelli$^2$, and S. Das Gupta$^3$}

\affiliation{$^1$Variable Energy Cyclotron Centre, 1/AF Bidhan Nagar, Kolkata700064,India}

\affiliation{$^2$LPC Caen IN2P3-CNRS/EnsiCaen et Universite, Caen, France}

\affiliation{$^3$Physics Department, McGill University, 
Montr{\'e}al, Canada H3A 2T8}

\date{\today}

\begin{abstract}

 Different formulas relying measurable fragment isotopic observables to the
 symmetry energy of excited nuclei have been proposed and applied to the analysis of heavy ion collision data in the recent literature. In this paper we examine the quality of the different expressions in the framework of the McGill Canonical Thermodynamic Model. We show that even in the idealized situation of canonical equilibrium and in the absence of secondary decay, these formulas do not give a precise reconstruction of the symmetry energy of the fragmenting source. 
However, both isotopic widths and isoscaling appear very well correlated to the physical symmetry energy. 
 
\end{abstract}


\maketitle
 
\section{Introduction}
 
One of the greatest challenge of intermediate energy heavy ion physics 
consists in the experimental measurement of the modifications to the symmetry energy $c_{sym}$ 
induced by density and temperature conditions different from the ones of ground state nuclei\cite{baoan}.
For the different astrophysical applications linked 
to the evolution and structure of compact stars\cite{skin,pierre}, 
the symmetry energy behavior at density far from saturation is of outmost importance.
The high density behavior of the isovector equation of state is almost not constrained by experimental observations, and considerable uncertainties exist also in the behavior of
the symmetry energy at subcritical density\cite{baran}. In this regime, nuclear matter is unstable with  respect to phase separation, mean-field estimation can become severely incorrect\cite{horowitz}, and clusterization has to be considered\cite{ropke,lehaut}.
Another important point concerns the temperature dependence of the symmetry energy, which is schematically treated or even neglected in the supernova explosion and proto-neutron star cooling modelizations\cite{horowitz,page}. 

A possible approach to this problem consists in comparing selected isospin observables to the output of a transport model where the isovector part of the equation of state can be varied\cite{msu}. This strategy has recently lead to very stringent constraints on the symmetry energy behavior\cite{tsang}, that appear reasonably consistent with the experimental results extracted from collective modes\cite{pigmy}, nuclear massses\cite{pawel} and neutron skins\cite{skins}.
The drawback of these analyses is that the results are model dependent, and different models do not produce fully consistent results\cite{baran,chimera}; moreover no information can be extracted about the finite temperature behavior.

Alternatively, a simple formula has been proposed\cite{botvina} to extract directly 
the symmetry energy  
from experimental cluster properties obtained
in the fragmentation of two systems 
of charge $Z_1,Z_2$, mass $A_1,A_2$ at the same temperature $T$:
\begin{equation}
4 \frac{c_{sym}^0}{T} = \frac{\alpha}{\left(Z_1^2/A_1^2\right)-\left(Z_2^2/A_2^2\right)}
\label{equBotvina} 
\end{equation}
where $\alpha$ is the so-called isoscaling parameter that can be 
measured from isotopic yields\cite{isoscaling}. 
Applications of this formula to different fragmentation data\cite{shetty,indra}
show that the measured symmetry energy tends to decrease with increasing collision
violence.

However eq.(\ref{equBotvina}) is not an exact expression. It was derived in the framework of macroscopic statistical models~\cite{botvina}, where many-body correlations are supposed to be entirely exhausted by clusterisation, and it appears to be strongly affected by conservation laws and combinatorial effects\cite{dasgupta,claudio};{ secondary decays may strongly affect the value of $\alpha$\cite{smf};}  finally, the $c_{sym}$ coefficient appearing 
 in eq.(\ref{equBotvina}) should correspond to the symmetry free-energy\cite{natowitz}, which is equivalent to the symmetry energy only in the $T\to 0$ limit.

In particular, S. Das Gupta et al. have shown that the difference in the neutron chemical potential $\Delta\mu_n$ increases with the temperature  if $c_{sym}$ 
is taken as a constant\cite{dasgupta}. In the grand-canonical ensemble, the difference of chemical potential between two sources at a common temperature $T$ is linked to the isoscaling parameter by 
\begin{equation}
\Delta\mu_n=\alpha T
\label{deltamu}
\end{equation}
This means that eq.(\ref{equBotvina}) cannot be exact at finite temperature. 
Experimentally the measured value of $\alpha T$ decreases with increasing incident energy and/or collision violence.
Let us suppose that the grand-canonical equality (\ref{deltamu}) 
is true also in the data, and
that ncreasing collision violence does indeed correspond to increasing temperature.
Then this implies that the physical symmetry energy coefficient explored in fragmentation data is not constant, but it is decreasing more steeply than obtained in the rough data\cite{shetty,indra}. 
This is consistent with the interpretation of ref.\cite{shetty,indra,botvina_csym,samaddar,baoan2}. But it is inconsistent with the statistical model calculations of ref.\cite{raduta}, where a constant input symmetry energy coefficient produces an apparent $c_{sym}$ 
from eq.(\ref{equBotvina}) qualitatively coherent with the experiment. 
The statistical model MMM used in ref.\cite{raduta} is similar, but not identical,
to the statistical model CTM used in ref.\cite{dasgupta}, which raises 
once again the question of the model dependence of the results.
However, refs.\cite{dasgupta},\cite{raduta} do not compute the same observables either, and use different prescriptions for the symmetry energy.
In these conditions, it is difficult to understand the real origin of the observed discrepancy. 

To progress on this issue, we present in this paper calculations with CTM 
made under similar conditions with respect to MMM. We will show that the two models produce 
very similar results.
Moreover, we address the question of an "improved" formula 
which would be valid out of the $T\to0$ limit. We will show that, in the framework of this model, none of the different formulas proposed in the literature allows a reliable direct measurement of the symmetry energy. However, both the isoscaling observable and isotopic widths appear very well correlated with the physical symmetry energy, implying that ratios of these isotopic observables measured in different systems should allow to extract the physical trend.

\section{The model}

The canonical thermodynamic McGill model is based on the analytic evaluation of
the canonical partition function for the fragmenting source with $A$ nucleons and $Z$
protons (neutron number $N=A-Z$) at a given temperature $T$ written as
\begin{eqnarray}
Q_{A,Z}=\sum\prod \frac{\omega_{a,j}^{n_{a,j}}}{n_{a,j}!}
\end{eqnarray}
where the sum is over all possible channels of break-up
which satisfy the conservation laws; $n_{a,j}$ is
the number of this composite in the given channel, and $\omega_{a,j}$ 
is the partition function of one composite with
nucleon number $a$ and proton number $j$:  
\begin{eqnarray}
\omega_{a,j}=\frac{V_f}{h^3}(2\pi maT)^{3/2}\times z_{a,j}(int)
\end{eqnarray}
Here $ma$ is the mass of the composite and
$V_f= V - V_0$ is the volume available for translational motion,
 where $V$ is the volume to which the system has expanded at break up and $V_0$ is the normal volume of $A$
nucleons and $Z$ protons.
Concerning  the choice of $z_{a,j}(int)$ used in this work,  the
proton and the neutron are fundamental building blocks 
thus $z_{1,0}(int)=z_{1,1}(int)=2$ 
where 2 takes care of the spin degeneracy.  For
deuteron, triton, $^3$He and $^4$He we use $z_{a,j}(int)=(2s_{a,j}+1)\exp(-
\beta e_{a,j}(gr))$ where $\beta=1/T, e_{a,j}(gr)$ is the ground state energy
of the composite and $(2s_{a,j}+1)$ is the experimental spin degeneracy
of the ground state.  Excited states for these very low mass
nuclei are not included.  For mass number $a=5$ and greater we use
the liquid-drop formula:
\begin{eqnarray}
z_{a,j}(int)=\exp\frac{1}{T}[W_0 a-\sigma(T)a^{2/3}-\kappa\frac{j^2}{a^{1/3}}
-c_{sym}\frac{(a-2j)^2}{a}+\frac{T^2a}{\epsilon_0}]
\label{partfunc}
\end{eqnarray}
The expression includes the 
volume energy, the temperature dependent surface energy, the Coulomb
energy, the symmetry energy and contribution from excited states in the continuum
since the composites are at a non-zero temperature. 
In this paper we will try different prescription for the symmetry energy coefficient, namely the same mass dependent prescription employed in the MMM model\cite{raduta}
\begin{equation}
c_{sym}=c_i c_v - c_i c_s a^{-1/3},
\label{csurf}
\end{equation}
with $c_i=1.7826$, $c_v=15.4941$ MeV, $c_s=17.9439$ MeV,
or alternatively a more sophisticated surface and temperature dependent expression\cite{das1}, accounting for the vanishing of all surface contributions at the critical point:
\begin{equation}
c_{sym}\left (a,T\right )=c_0-c_i T_c a^{-1/3}\left (\frac{T_c^2-T^2}{T_c^2+T^2} \right )^{5/4}.
\label{ctempsurf}
\end{equation}
where $c_0=28.165$ MeV, and $T_c=18$ MeV is the critical temperature.
To test the sensitivity of the different observables to the symmetry energy, a schematic constant coefficient will also be used. 

In using the thermodynamic model one needs to specify which composites
are allowed in the channels.  For mass numbers $a$=5 and 6, we include proton
numbers 2 and 3 and for mass number $a$=7, we include proton numbers 2,3 and 4. For $a\ge 8$, we include all nuclei within drip-lines
defined by the liquid-drop formula.

The Coulomb interaction between different composites is included in
the Wigner-Seitz approximation\cite{das1,Bondorf1}.   
For further details, see ref.\cite{das1}.
 
\section{Symmetry energy evaluations}

In this section we present the results of calculations made with the CTM model
with the aim of reconstructing the input symmetry energy of the model from measurable cluster observables. When not explicitely stated, we will consider an excited fragmented source composed of $Z=75$ protons with two different mass numbers $A=168$, $A=186$. This specific choice of source size was already employed in  previous works\cite{dasgupta,gargi}.
Table \ref{tab1} gives the value of the isoscaling parameter obtained in the model and the resulting apparent symmetry energy from eq.(\ref{equBotvina}) for different values of the temperature and the  break-up volume. The isoscaling parameter $\alpha$ is the value extracted from the slopes of differential cluster yields\cite{dasgupta} averaged over $Z=1,2,3,4,5$, similar
to the procedure employed in the analysis of heavy ion data\cite{msu}. For these light isotopes, an excellent isoscaling is observed in the model\cite{dasgupta}. 
The input symmetry energy in this exploratory calculation is fixed to $c_{sym}=23.5$ MeV.
\begin{table}[!htbp]
  \begin{center}
    \begin{tabular}{r|c|c|c|c}
       & Temperature & $V/V_0$ & $\alpha$ & $c_{sym}^0$  \\
      \hline
        & 7.5 $MeV$ & 4 & 0.511 & 26.10 $MeV$  \\
        & 6.5 $MeV$ & 4 & 0.557 & 24.65 $MeV$  \\
      & 5.5 $MeV$ & 4 &  0.606 &  22.72 $MeV$ \\
      & 4.5 $MeV$ & 4 &  0.703 &  21.56 $MeV$ \\
      & 3.5 $MeV$ & 4 &  0.870 &  20.75 $MeV$ \\
      
      & 7.5 $MeV$ & 6 &  0.462 &  23.6 $MeV$ \\
      & 6.5 $MeV$ & 6 &  0.514 &  22.75 $MeV$ \\
      & 5.5 $MeV$ & 6 &  0.578 &  21.65 $MeV$ \\
      & 4.5 $MeV$ & 6 &  0.673 &  20.63 $MeV$ \\
      & 3.5 $MeV$ & 6 &  0.923 &  22.02 $MeV$ \\
      
    \end{tabular}
  \end{center}
  \caption{ Isoscaling parameter averaged over $Z=1-5$ and apparent symmetry energy from eq.(\ref{equBotvina}) for a fragmenting source with $Z=75$ at different temperatures and break-up volumes. The input symmetry energy in the model is 
  $C_{sym}=23.5$ MeV independent of the temperature.
 }
  \label{tab1}
\end{table}

The isoscaling parameter $\alpha$ decreases with increasing temperature independent of the break-up volume. This is in agreement with the results of previous works \cite{raduta,dasgupta}, as well as with the experimental observation of a decreasing $\alpha$ with increasing collision violence. Concerning the extracted values of $c_{sym}$, an important dependence on the break-up volume is observed. Fot small break-up volumes, the apparent $c_{sym}$ monotonically increases with temperature
as already observed in a preceeding study with the CTM model where $\alpha$ was deduced from the chemical potential via the grancanonical expression eq.(\ref{deltamu})
and not deduced from the slope of fragment yields\cite{dasgupta}.
For higher volumes, the apparent $c_{sym}$ initially decreases as in ref.\cite{raduta} and in the data, and it increases again when $\alpha$ saturates. This second regime may not be explored in the data because the break-up temperatures saturate with increasing collision violence\cite{shetty}. The same may be true for the analysis of ref.\cite{raduta} which was made in the microcanonical ensemble; indeed in this latter the temperature does not increase linearly with excitation energy, producing a saturation in the apparent $c_{sym}$. The existence of these different behaviors
shows that grancanonical formulas have to be handled with care: in the canonical 
or microcanonical models the slope of isotopic yields may not be directly related to the chemical potential.

For the higher break-up volume, the input symmetry energy coefficient is well
recovered at the lowest temperature. This was expected since eq.(\ref{equBotvina}) has been derived in the limit of vanishing temperature. 
Surprisingly, this does not seem the case if the break-up volume is small. In this case, the limit may be attained at lower temperatures where our fragmentation model cannot be safely applied any more. However,
for all the situations considered, which cover a large range of thermodynamic conditions typically accessed in fragmentation experiments, the deviation between the input $c_{sym}$ and the approximation extracted from eq.(\ref{equBotvina}) never exceeds $12\%$, which is a reasonable precision considering the inevitable error bars 
induced by efficiency, event selection and thermometry in heavy ion collision experiments.

\begin{figure}[htbp]
\includegraphics[width=5.5in,height=4.5in,clip]{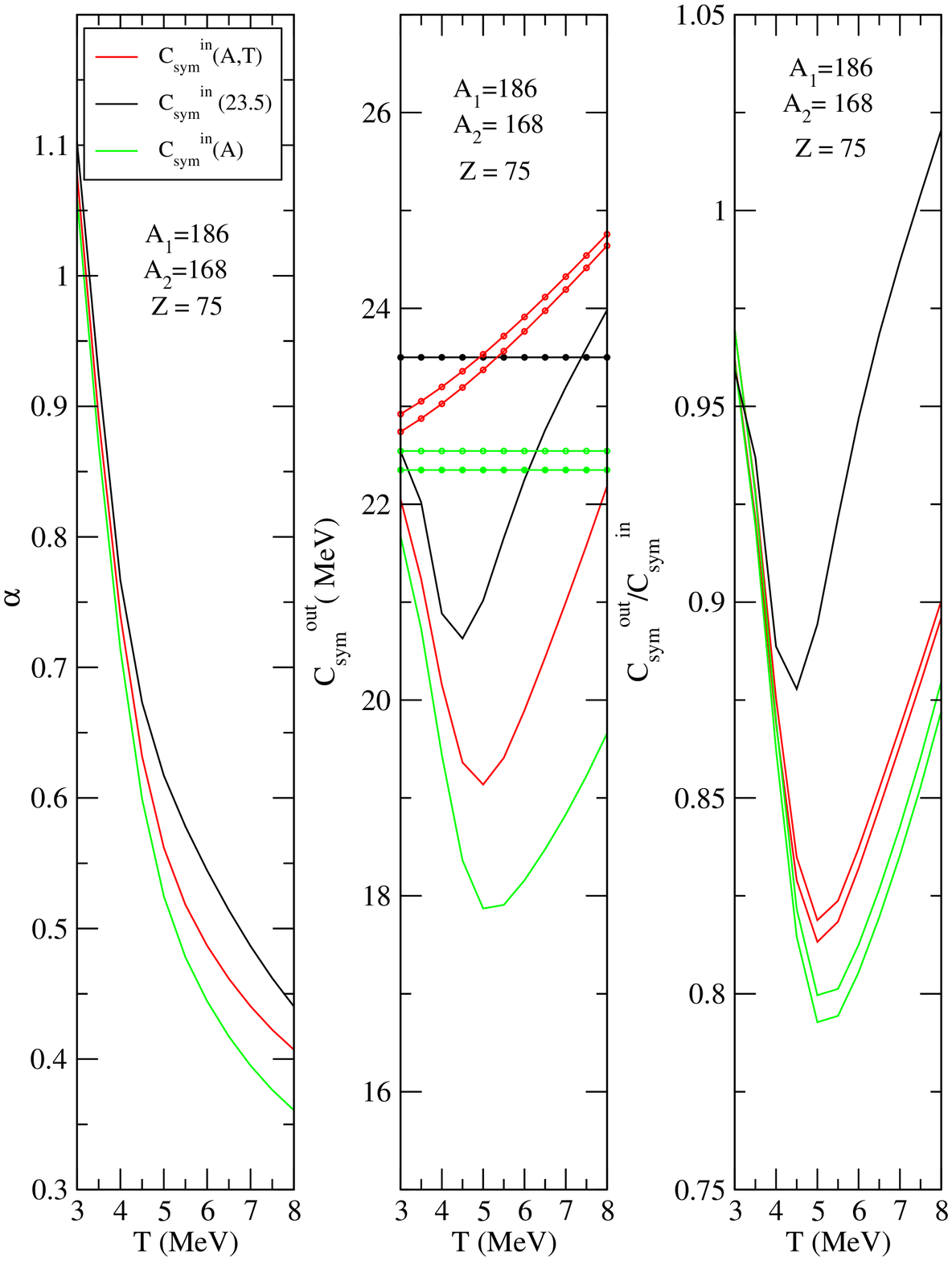}
\caption{ As a function of the temperature, the left side shows the isoscaling $\alpha$ parameter, the middle part shows the resulting apparent $c_{sym}$ from eq.(\ref{equBotvina}), and the right part the ratio between the apparent $c_{sym}$
and the input $c_{sym}$ of the source. Upper lines (black online): constant $c_{sym}=23.5$ MeV. Middle lines (red online): mass and temperature dependent $c_{sym}$ from
eq.(\ref{ctempsurf}).
Lower lines (green online): mass dependent $c_{sym}$ from eq.(\ref{csurf}). The lines with symbols in the middle panel give the input symmetry energy for a mass equal to the source(s) mass.
The freeze-out volume is fixed to $V/V_0=6$ }
\label{fig1}
\end{figure}

To further progress on this analysis, we plot on the left part of Fig.\ref{fig1} the isoscaling $\alpha$ parameter as a function of the temperature with different choices of the symmetry energy parametrization. 
In all cases a decreasing isoscaling parameter is found.
The middle part of the same figure shows the resulting symmetry energy coefficient obtained by applying eq.(\ref{equBotvina}). 
We can see that the functional form of $c_{sym}$ does not affect strongly the trend of the results. In particular, a decreasing $\alpha$ does  not necessarily imply a decreasing physical symmetry energy. Moreover, the temperature and surface dependence of the physical symmetry energy affects the predictive power of eq.(\ref{equBotvina}).
In no case the extracted coefficient approaches the symmetry energy of the fragments
used for the isoscaling analysis, however at a given value of temperature, it qualitatively follows the trend of the input symmetry energy of the fragmenting source, as expected in the Weisskopf regime\cite{weisskopf}. 
This means that the isoscaling properties of the lightest fragments appear well correlated to the symmetry energy of their emitting source, even out of the evaporation regime. 

From the observations of Fig.\ref{fig1} and table \ref{tab1} we can already draw some partial conclusions.  An important point concerns the fact that observing $\alpha$  or $c_{sym}^0$ decreasing with the collision violence cannot be taken as an evidence that the physical symmetry energy does so.  Only a detailed comparison with a model 
may allow to extract the physical symmetry energy.   
Different models have to be very carefully compared to a large set of independent observables before one can extract any conclusion.
As a second remark,  both MMM and CTM models tend to agree that at low temperature eq.(\ref{equBotvina}) gives a good reproduction of the physical symmetry energy. 
 
This means that results obtained from intermediate impact parameter collisions in the neck region\cite{rizzo,casini} (where in principle the matter is at low density but also relatively cold),
or analyses of quasi-projectiles produced in peripheral collisions\cite{shetty} (where the system is close to normal density and the temperature behavior could be disentangled from the density behavior) are better suited to this study than central collisions in the multifragmentation regime.

To progress on the issue of the determination of the in-medium modifications
to the symmetry energy, it would be extremely useful if we could have a formula more adapted to the finite temperature case. 
To this purpose, we turn to check two other expressions proposed in the literature 
to access the symmetry energy from fragment observables.

\section{The influence of fractionation}

In nuclear multifragmentation reactions, the asymmetry term influences
the neutron-proton composition of the break-up fragments.

Interpreting multifragmentation in the light of first-order phase transitions
in multicomponent systems,
the neutron enrichment of the gas phase with respect to the liquid phase
comes out as a natural consequence of Gibbs equilibrium criteria
and a connection between phases chemical composition and the symmetry term
can be established \cite{mueller,gulminelli}.
Interesting enough, the phenomenon of isospin fractionation which is 
systematically observed in analyses of multifragmentation 
data \cite{xu,geraci,martin,shetty,botvina}, seems to be a generic feature of phase
separation independent of the equilibrium Gibbs construction \cite{isospinfrac}. 
Indeed, dynamical models of heavy ion collisions \cite{baoan,baran}
where fragment formation is essentially ruled by the out of equilibrium process of 
spinodal decomposition also exhibit fractionation.
Adopting an equilibrium scenario for the break-up stage of a multifragmenting system,
Ono {\it et al.} \cite{ono} derive an approximate grandcanonical  expression which connects
the symmetry term with the isotopic composition of 
fragments obtained in the break-up stage of two 
sources with similar sizes in identical thermodynamical states 
and differing in their isospin content,
\begin{equation}
c_{sym}(j)=\frac{\alpha(j) T}{4 \left[ \left( \frac{j}{<a>_1}\right)^2-
\left( \frac{j}{<a>_2}\right)^2\right] },
\label{eq:csym_ono}
\end{equation}
under the hypothesis that the isotopic distributions are essentially Gaussian
and that the free energies contain only bulk terms.
Here, $\alpha(j)$ is the isoscaling slope parameter of a fragment of charge $j$,
and $<a>_i$ stands for the average mass number of a fragment of charge $j$ produced
by the source $i(=1,2)$ at the temperature $T$. 

In the limit of vanishing temperature, fractionation can be neglected
and $j/<a>_i$ can be replaced by the corresponding quantity of the sources $Z_i/A_i$ \cite{botvina} giving back eq.(\ref{equBotvina}). 
In the opposite case, fragment yields are predicted to be sensitive to their proper symmetry energy and not to the symmetry energy of the emitting source.

\begin{figure}[htbp]
\includegraphics[width=5.5in,height=4.5in,clip]{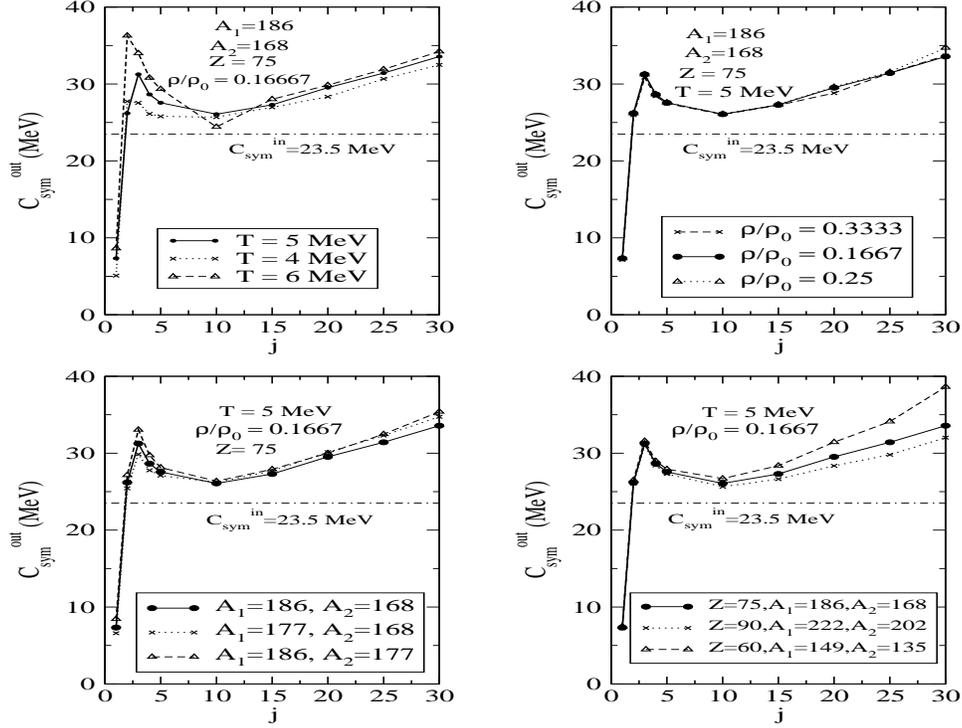}
\caption{Apparent symmetry energy calculated from eq.(\ref{eq:csym_ono}) as a function of the cluster charge. Upper left: calculations at different temperatures $T=4$ MeV (dotted
line), $T=5$ MeV (full line), $T=6$ MeV (dashed line) for a source charge $Z=75$, source masses $A_1=168$, $A_2=186$, and a volume $V/V_0=6$.
Upper right: calculations at different volumes $V/V_0=3$ (dashed line), $V/V_0=4$  (dotted line), $V/V_0=6$ (full line) for a source charge $Z=75$, source masses $A_1=168$, $A_2=186$, and a temperature $T=5$ MeV.
Lower right: calculations at different isospin values $A_1=177$ and $A_2=168$ (dotted line),  $A_1=186$ and $A_2=168$(full line),  $A_1=186$ and $A_2=177$ (dashed line) for a source charge $Z=75$, a temperature $T=5$ MeV, and a volume $V/V_0=6$.
Lower left: calculations for different source sizes $Z=75$, $A_1=186$ and $A_2=168$ (full line), $Z=60$, $A_1=149$ and $A_2=135$(dashed line),  $Z=90$, $A_1=222$ and $A_2=202$ (dotted line) for a source temperature $T=5$ MeV, and a volume $V/V_0=6$.
  }\label{fig2}
\end{figure}

Figure \ref{fig2} gives, as a function of the cluster atomic number $j$, the apparent symmetry energy coefficient extracted from eq.(\ref{eq:csym_ono}) for different conditions of temperature, free volume, source isotopic content, and source size.
In all cases the input symmetry energy has been fixed to the constant value $c_{sym}=23.5$ MeV. We can see that eq.(\ref{eq:csym_ono}) leads to a global systematic 
overestimation of the input symmetry energy. 
The response of the different fragments depends on their size. Non realistic values
are obtained for the lightest fragments.
The lightest fragments isotopic distribution $j/<a>$ is very sensitive to the number of isotopes considered in the calculation, which show strong binding energy fluctuations. It is not surprising that they cannot be treated by eq.(\ref{eq:csym_ono}), which implies a behavior dominated by the bulk for all fragments. 
Turning to the increase in the apparent symmetry energy for the heaviest fragments, this  is most probably due to the failing of the grancanonical approximation in eq.(\ref{eq:csym_ono}) when the fragment size becomes comparable to the source size, as previously discussed in ref.\cite{raduta}. One should also note that for heavy fragments isoscaling tends to be violated in the model\cite{dasgupta}, and the determination of an isoscaling slope becomes largely arbitrary. Clusters of charge $j>5$ and smaller than approximately one tenth of the source size are best suited to this analysis. If we limit ourselves to such intermediate mass fragments, we can see that the apparent $c_{sym}$ coefficient is reasonably independent of the available volume, source isospin and mass. A temperature dependence is still apparent and, as in the case of Fig.\ref{fig1}, does not exceed 10\%. These results are in good agreement with the findings of ref.\cite{raduta} in the framework of the MMM model.

\begin{figure}[htbp]
\includegraphics[width=5.5in,height=4.5in,clip]{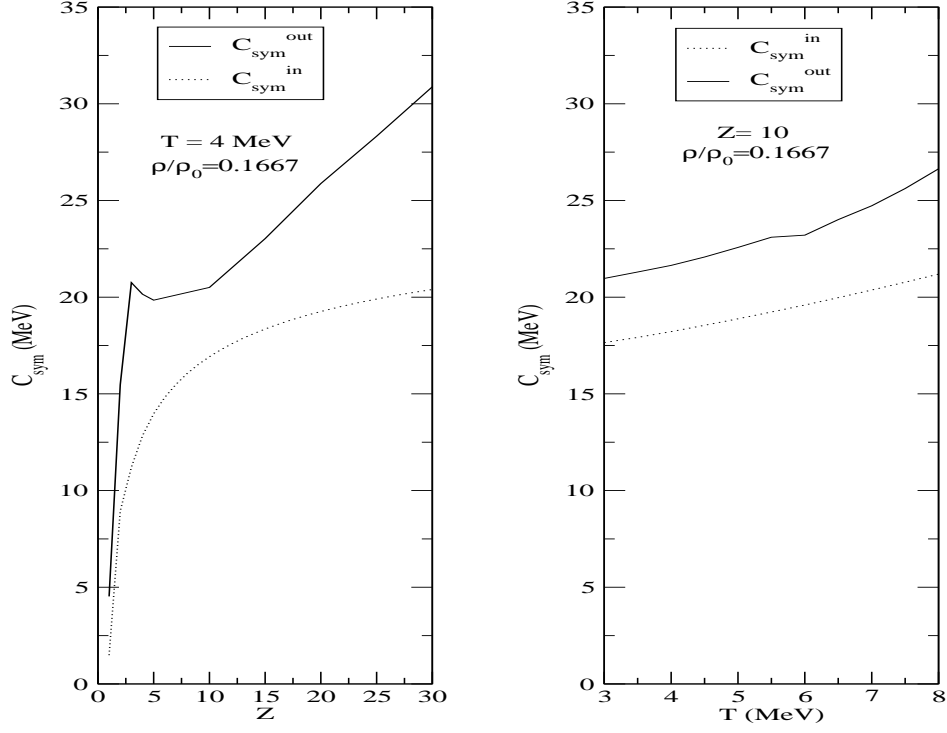}
\caption{Left side: Apparent symmetry energy calculated from eq.(\ref{eq:csym_ono}) as a function of the cluster charge for a volume $V/V_0=6$. Left side:  Temperature $T=4$ MeV
and the input $c_{sym}$ displayed by dotted line is taken from eq.(\ref{csurf}).  Right side: same as left, but for a fragment charge $Z=10$ as a function of the temperature
and the input $c_{sym}$ displayed by dotted line is taken from eq.(\ref{ctempsurf}). 
 }
\label{fig3}
\end{figure}

Figure \ref{fig3} shows the response of eq.(\ref{eq:csym_ono}) to a symmetry energy depending on the temperature and on the fragment size through eq.(\ref{ctempsurf}).
The behavior is very similar to the one displayed in Figure \ref{fig2} above.
This means that it is not possible to extract the surface dependence by looking at the behavior as a function of the charge. The temperature dependence for a fixed charge (right part of Figure \ref{fig3}) conversely shows a good correlation, meaning that the temperature dependence could be extracted studying the isoscaling for a given charge at different excitation energies.

An alternative expression has been derived in ref.\cite{raduta2} connecting
the symmetry energy of a cluster of size $a$ to the width of its isotopic distribution.
Indeed, a Gaussian approximation on the grandcanonical expression for cluster yields
gives
\begin{equation}
\sigma_I^2(a) \approx \frac{aT}{2c_{sym}(a)},
\label{eq:csym_fluct_fr}
\end{equation}
where $\sigma_I^2(a)$ indicates the width of the isotopic distribution of a cluster of size $a$, and $I=a-2j$.
In principle the $c_{sym}$ coeffcients appearing in eqs.(\ref{eq:csym_ono}) and  
(\ref{eq:csym_fluct_fr}) correspond to symmetry free-energy coefficients, that is
they include an entropic contribution.
However if we neglect the $I$ dependence of the excitation energy and entropy
associated to a given mass $a$, they coincide with $c_{sym}$ defined by eq.(\ref{partfunc}).

Fig.\ref{fig3} shows the apparent $c_{sym}$ extracted from the fluctuation formula 
eq.(\ref{eq:csym_fluct_fr}) as a function of the cluster charge $j$ at different temperatures and isospin
values. This observable shows a linearly increasing behavior similar to the one 
displayed by eq.(\ref{eq:csym_ono}), but the overall quality of reproduction is improved.

In this picture the input symmetry energy was taken as a constant. To see if eqs.(\ref{eq:csym_ono}),(\ref{eq:csym_fluct_fr}) can be used to extract from observable cluster data the possible surface/temperature/density dependence of 
the symmetry energy, we additionally show in Figure \ref{fig5}, for a cluster charge $j=10$,
the apparent $c_{sym}$ extracted from eqs.(\ref{eq:csym_ono}),(\ref{eq:csym_fluct_fr})  as a function of the input $c_{sym}$. We can see that both formulas
produce a bias. This implies that a quantitative estimation of the symmetry energy 
coefficient cannot be obtained from these expressions. We recall that 
the presented calculation completely neglects secondary decay, which is expected to considerably increase this bias. 

If the quantitative values do not match, we observe however a good linear
correlation in both cases.
This means that the analyzed observables show an excellent sensitivity to the isovector equation of state. In particular, the relative variation
of the extracted symmetry coefficient should be reliable, and it would be very important to check whether the linear dependence survives to secondary decay.
We recall that in the framework of the MMM model which (at variance with CTM) 
contains an afterburner, the proportionality is kept also after secondary decay\cite{raduta2}.

\section{Conclusions}

In this paper we have studied the sensitivity to the symmetry energy of different isotopic observables measurable in heavy ion collisions, in the framework of the McGill Canonical Thermodynamic Model.

We conclude that, even in the idealized limit of thermal equilibrium, no isotopic observable can allow to reconstruct  the physical symmetry energy of the excited fragmenting source in a model indpendent way. The different models have to be very carefully compared to a large set of independent observables before one can extract any conclusion. 
In the low temperature limit, the well spread expression eq.(\ref{equBotvina})
gives a good measurement of the symmetry energy of the source, and is never sensitive to the symmetry energy of the fragments. At high temperature in the multifragmentation regime, no formula gives a satisfactorily reproduction of the input $c_{sym}$.
 
However we confirm, in agreement with the results of previous studies with different models\cite{dasgupta,raduta,raduta2,shetty,msu,tsang} that both the isoscaling variable and isotopic widths show a very strong sensitivity to the strength of the symmetry energy.

This means that the use of the different formulas proposed in the literature should allow to better constrain $c_{sym}$. 
In particular, if a drastic reduction of the effective $c_{sym}$ 
in the nuclear medium is observed, we should be able to see it 
by calculating the differential response of eqs.(\ref{eq:csym_ono}),(\ref{eq:csym_fluct_fr}) 
for different excitation energies.

It is important to stress that this study neglects secondary decay which may 
have dramatic consequences on isotopic observable\cite{smf}. 
It is clear that the issue should be investigated. Moreover differential observables\cite{msu,tsang} where the effect of secondary decay may 
cancel out should be analysed.

\begin{figure}[htbp]
\includegraphics[width=5.5in,height=4.5in,clip]{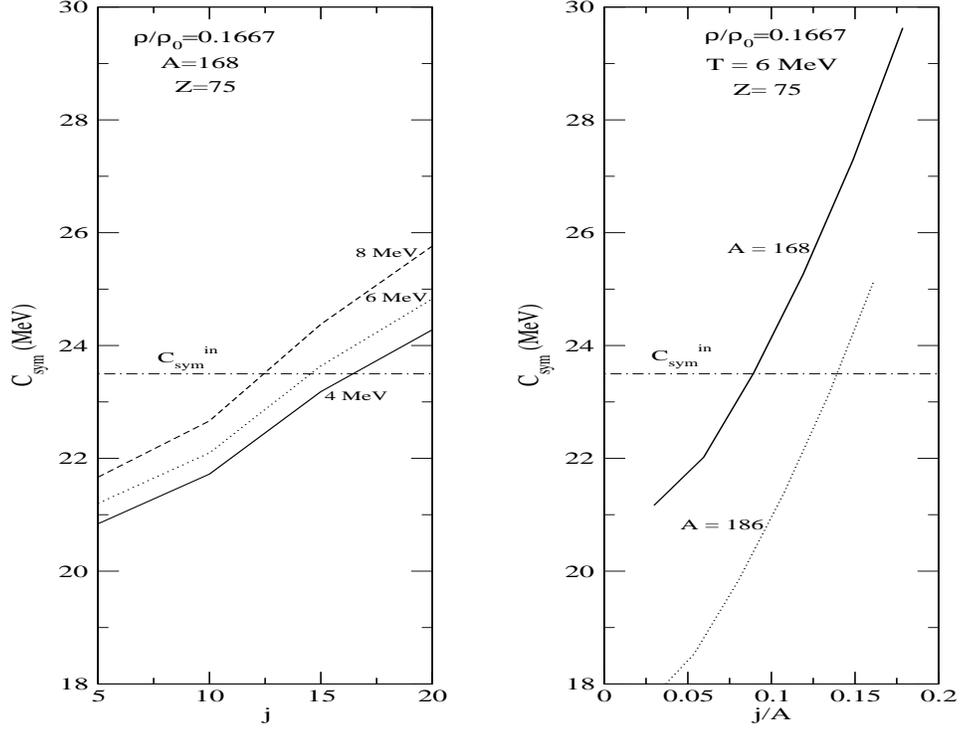}
\caption{Left side: apparent $c_{sym}$ extracted from the fluctuation formula 
eq.(\ref{eq:csym_fluct_fr}) as a function of the cluster charge $j$
for a source charge $Z=75$, source mass $A_1=168$, and a volume $V/V_0=6$, and temperature $T=4$ MeV (full line), $T=6$ MeV (dotted line), $T=8$ MeV (dashed line). Right side: same as left, but for a temperature $T=6$ MeV and source mass $A_1=168$
(full line) and $A=186$ (dotted line). The x axis is $j/A$.
 }
\label{fig4}
\end{figure}

\begin{figure}[htbp]
\includegraphics[width=5.5in,height=4.5in,clip]{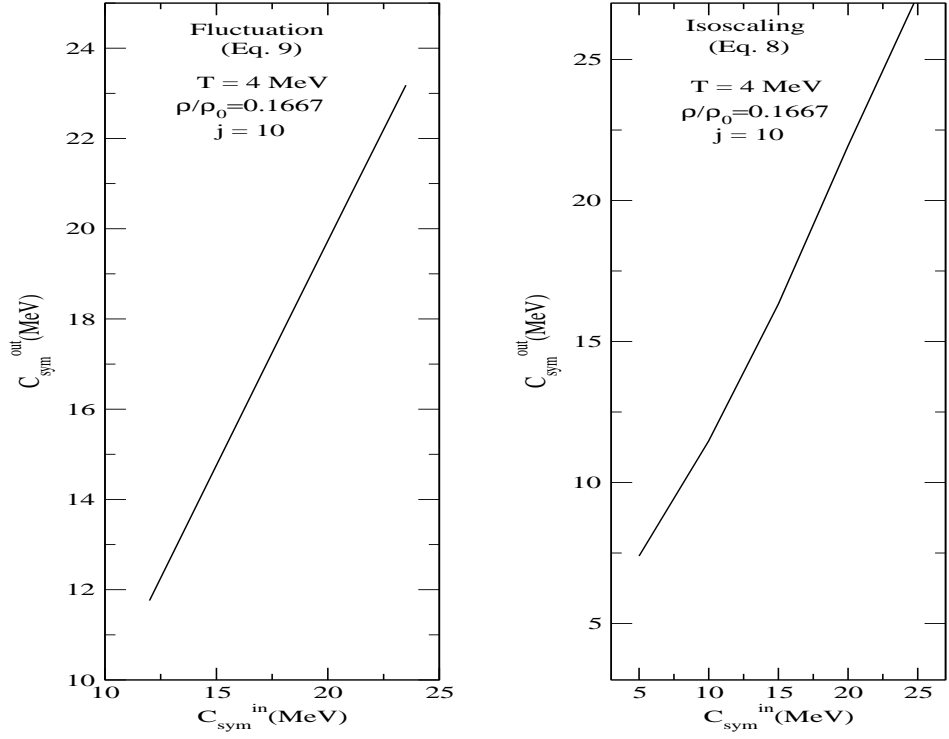}
\caption{apparent $c_{sym}$ extracted from eq.(\ref{eq:csym_fluct_fr}) (left side) and eq.(\ref{eq:csym_ono}) (right side)  as a function of the input $c_{sym}$  for a cluster
charge $j=10$. Conditions are as in Figs.\ref{fig2},\ref{fig4} above.}\label{fig5}
\end{figure}

\end{document}